\documentclass[reprint,
 amsmath,amssymb,
 aps,
]{revtex4-1}

\usepackage{xcolor}
\usepackage{graphicx}
\usepackage{dcolumn}
\usepackage{bm}
\usepackage{tensor} 
\begin{document}
\title{Early-time thermalization of cosmic components? A hint for solving cosmic tensions}

\author{Hermano Velten$^{1}$}%
 \altaffiliation[]{hermano.velten@ufop.edu.br (corresponding author)}
 \author{Ingrid Costa$^{2}$}
 \author{Winfried Zimdahl$^{2}$}
\affiliation{%
$^{1}$Departamento de F\'isica, Universidade Federal de Ouro Preto (UFOP), Campus Universit\'ario Morro do Cruzeiro, 35.400-000, Ouro Preto, Brazil}
\affiliation{%
$^{2}$N\'ucleo COSMO-UFES \& Departamento de F\'isica,  Universidade Federal do Esp\'irito Santo (UFES)\\
 Av. Fernando Ferrari s/n CEP 29.075-910, Vit\'oria, ES, Brazil }%
\date{\today}

\begin{abstract}
We study an expanding two-fluid model of non-relativistic dark matter and radiation which are allowed to interact during a certain time span and to establish an approximate thermal equilibrium.
Such interaction which generates an effective bulk viscous pressure at background level is expected to be relevant for times around the transition from radiation to matter dominance. 
We quantify the magnitude of this pressure for dark matter particles masses within the range $1 {\rm eV} \lesssim m_{\chi} \lesssim 10 {\rm eV}$ around the matter-radiation equality epoch (i.e., redshift $z_{\rm eq}\sim 3400$) and demonstrate that the existence of a transient bulk viscosity has consequences which may be relevant for addressing current tensions of the standard cosmological model: i) the additional (negative) pressure contribution modifies the expansion rate around $z_{\rm eq}$, yielding a larger $H_0$ value and ii) large-scale structure formation is impacted by suppressing the amplitude of matter overdensity growth via a new viscous friction-term contribution to the Mészáros effect. 
As a result, both the $H_0$ and the $S_8$ tensions of the current standard cosmological model are significantly alleviated. 
\end{abstract}

\maketitle

\section{Introduction} 

The cosmological large-scale structure (LSS) seen today is the final outcome of a process that started during the primordial inflationary universe which sets the initial conditions for the density field, followed by the posterior standard-model evolution. 
At the early stages of structure formation, the radiation-dominated background impedes the efficient growth of sub-horizon density fluctuations by speeding up the background evolution. When matter becomes the protagonist as the background expansion driver, the expansion slows down and clustering is favored. According to the standard cosmological model this happens around a redshift $z_{\rm eq}\sim 3400$, a moment known as the matter-radiation equality epoch i.e., when both radiation and matter energy densities are the same $\rho_r=\rho_m$. The change in the background expansion is the physical mechanism behind the so called Mészáros effect \cite{Meszaros:1974tb}. 

The background expansion transition also leaves imprints on super-horizon modes. A widely known result is related to the fact that the gravitational potential amplitude for scales larger than the horizon is reduced by $10\%$ across the transition from radiation to matter dominated epochs. This has important consequences later on for the Cosmic Microwave Background (CMB) temperature distribution \cite{amendola, mo}.

Internally, the radiation fluid is a baryon-photon plasma in which photons can efficiently transfer energy from different regions of the fluid via e.g., a diffusion mechanism. Also, at perturbative level shear viscosity and heat conduction may play a role, leading to the Silk damping effect \cite{Silk}. 

Apart from few attempts in Refs. \cite{Pan:2018zha,Das:2017nub},
the possibility that radiation and dark matter components can interact at times before recombination  is still poorly explored so far.

The attempt in Ref. \cite{Zimdahl:1996fj} is based on a Friedmann-Lema\^itre-Robertson-Walker (FLRW) cosmological evolution of two adiabatic fluids that are allowed to establish thermal equilibrium (defined by an equilibrium temperature $T$ of the system as a whole). Fluid particles are interacting so weakly that the energy of their interaction may be neglected and one can assume that the total energy density of the composite gas is the sum of the energy densities of the components. On the other hand, this interaction is strong enough to maintain an approximate 
 equilibrium with an only small non-equilibrium contribution.

It has been demonstrated in \cite{Zimdahl:1996fj} that such small non-equilibrium contributions can be mapped onto an effective bulk viscosity of the system as a whole.

This result exemplifies the widely known fact that multi-fluid systems are intrinsically non-adiabatic. While dissipative physics is a customary feature in modelling the dynamics of real fluids such aspects are not present in the standard cosmological model. However, the phenomenology associated to cosmological bulk viscous models is quite abundant in the literature \cite{Zimdahl:1996ka, Colistete:2007xi,Brevik:2020psp,Normann:2016jns,Szydlowski:2020awp, Brevik:2017msy,Barbosa:2015ndx}. 

In the next section we review the main results of Ref. \cite{Zimdahl:1996fj} and apply them to a two-fluid system of matter and radiation. The emerging transient bulk viscous pressure acts as an extra (and new) background effect during the radiation-matter transition epoch, disappearing both in the early and future time limits. We quantify the magnitude of such new background bulk viscous pressure and compute in detail its consequences for the LSS  evolution, namely, the impact on the Mészáros effect and on the evolution of the super-horizon gravitational potential around $z_{\rm eq}$.

\section{Effective (one-fluid) viscous dynamics from two coupled perfect fluids}

Let us start considering that the total energy momentum tensor of the cosmic medium can be written as the sum of two components as
\begin{equation}
    T^{\mu\nu}=T^{\mu\nu}_1 + T^{\mu\nu}_2.
\end{equation}

Individually, each component $A= 1, 2$ has the perfect fluid structure $T^{\mu\nu}_A=(\rho_A+p_A) u^{\mu}u^{\nu}+p_A g^{\mu\nu}$ and obeys the energy density and particle number density conservation laws, respectively
\begin{eqnarray}
&&T^{\mu\nu}_{A;\nu}=\dot{\rho}_A+\Theta(\rho_A + p_A)=0, \\ \nonumber
&&   N^{\nu}_{A;\nu}=\dot{n}_A+\Theta n_A=0,
\label{consrhoN}
\end{eqnarray}
where $\Theta=u^{\mu}_{; \mu}$ is the expansion scalar, $N^{\nu}_{A}= n_Au^{\nu}$ is the particle-number flow vector of component A and $n_A$ is the corresponding particle-number density.  In a FLRW universe $\Theta=3H$.
Allowing for a thermal interaction between both components one can implement an effective one-fluid description of this system by defining global quantities like the total particle number density $n=n_1+n_2$, the overall pressure $p(n,T)$ and {\ the total energy density $\rho(n,T)$.} Following Ref. \cite{Udey}, the equilibrium temperature $T$ is defined by the relation 
\begin{equation}
    \rho_1(n_1,T_1)+\rho_2(n_2,T_2)=\rho(n,T).
\label{energycons}
\end{equation}
Particle number densities and temperatures have been taken here as the basic thermodynamical variables.
As shown in \cite{Zimdahl:1996fj}, the above condition implies that $p_1(n_1,T_1)+p_2(n_2,T_2) \neq p(n,T)$. This difference is associated to the emergence of a bulk viscous pressure $\Pi$ that is defined accordingly
\begin{equation}
    \Pi= p_1(n_1,T_1)+p_2(n_2,T_2) - p(n,T).
    \label{Pressurebalance}
\end{equation}
The system is assumed to be at equilibrium at a certain time $\eta_0$, then $T(\eta_0)= T_1(\eta_0)= T_2(\eta_0)$, $p(\eta_0)=p_1(\eta_0)+p_2(\eta_0)$. 
During a subsequent time interval $\tau$, each component follows its own internal perfect fluid dynamics such that  at a time $\eta_0 + \tau$ 
up to first order 
\begin{eqnarray}
\rho_A(\eta_0 +\tau)=\rho_A(\eta)+\tau \dot{\rho}_A + ... \\ \nonumber
\rho(\eta_0 +\tau)=\rho(\eta)+\tau \dot{\rho} + ...
\end{eqnarray}
is valid.

For different equations of state the equilibrium temperatures of the (perfect) fluids evolve differently, 
\begin{equation}\label{dotTA}
    \dot{T}_A(\eta_0)=- 3H T_A \frac{\partial p_A/ \partial T_A }{\partial \rho_A/ \partial T_A} .
\end{equation}
Here, the partial derivatives with respect to the temperatures have to be taken at fixed particle number densities. 
For the overall equilibrium temperature one has at $\eta_0$, 

\begin{equation}\label{dotT}
    \dot{T}(\eta_0)=- 3H T \frac{\partial p/ \partial T }{\partial \rho/ \partial T } .
\end{equation}
It follows that at a time $\eta_0 + \tau$ there appear first-order temperature differences 
\begin{eqnarray}
&&T_1-T_2=-3H \tau  T \left(\frac{\partial p_1/\partial T}{\partial \rho_1/\partial T }-\frac{\partial p_2/\partial T}{\partial \rho_2/\partial T }\right), \\
&&T_1-T=- 3H\tau  T \left(\frac{\partial p_1/\partial T}{\partial \rho_1/\partial T }-\frac{\partial p/\partial T}{\partial \rho/\partial T }\right), \\
&&T_2-T=-3H \tau a T \left(\frac{\partial p_2/\partial T}{\partial \rho_2/\partial T }-\frac{\partial p/\partial T}{\partial \rho/\partial T }\right).
\end{eqnarray}
These differences are the result of the different cooling rates of the individual components during the time interval $\tau$. 

At the instant $\eta_0+\tau$ one has $T(\eta_0+\tau)\neq T_1(\eta_0+\tau)\neq T_2(\eta_0+\tau)$ and the sum of the partial pressures reads
\begin{eqnarray}
    p_1(n_1,T_1)+p_2(n_2,T_2)=p_1(n_1,T)+p_2(n_2,T) \\ \nonumber
    +(T_1-T)\frac{\partial p_1}{\partial T}+(T_2-T)\frac{\partial p_2}{\partial T}.
\end{eqnarray}
The temperature difference terms give rise to a bulk viscous pressure $\Pi$, i.e., 
\begin{equation}\label{totalpressure}
    p_1(n_1,T_1)+p_2(n_2,T_2)=p(n,T)+\Pi.
\end{equation} 
According to the Eckart theory \cite{Eckart} for dissipative fluids, i.e., $\Pi=-3H \xi$, the bulk viscous coefficient $\xi$ is given by
\begin{equation}
    \xi = - \tau T \frac{\partial \rho}{\partial T}\left(\frac{\partial p_1}{\partial \rho_1}-\frac{\partial p}{\partial \rho}\right)\left(\frac{\partial p_2}{\partial \rho_2}-\frac{\partial p}{\partial \rho}\right).
    \label{xi}
\end{equation}
Here,
\begin{equation}
\frac{\partial p_A}{\partial \rho_A} \equiv \frac{\partial p_A/\partial T_A}{\partial \rho_A/\partial T_A}, \quad 
\frac{\partial p}{\partial \rho} \equiv \frac{\partial p/\partial T}{\partial \rho/\partial T}, 
    \label{}
\end{equation}
i.e., in all partial derivatives the number densities have to be kept fixed. 
Formula (\ref{xi}) for the bulk viscous coefficient $\xi$ is the main result of Ref. \cite{Zimdahl:1996fj}. 

Let us apply now (\ref{xi}) to a system radiation and dark matter. Let us identify fluid $1$ with radiation i.e., $p_1=p_r$ and fluid $2$ with a scalar dark matter particle $\chi$, $p_2=p_{\chi}$. Our aim is find an expression for the effective bulk viscosity of the mixture radiation-matter which could have been relevant at the radiation-matter transition epoch. 

Since we are formulating the dynamical description of cosmic fluids using the particle number density $n$ and the temperature $T$ as basic thermodynamical variables the relevant equations of state read
\begin{eqnarray}
    &&p_r=n_r k_B T_r, \quad \rho_r=3 n_r k_B T_r, \\
    &&p_{\chi}=n_{\chi} k_B T_{\chi},  \quad \rho_{\chi} = n_{\chi} m_{\chi}c^2 + \frac{3}{2}n_{\chi} k_B T_{\chi},
\label{EoS}\end{eqnarray}
where $k_B$ is the Boltzmann constant and $m_{\chi}$ is the mass of the dark matter particle. Using these equations of state 
one finds (cf. Eq (53) in \cite{Zimdahl:1996fj}),
\begin{equation}\label{xi+}
    \xi=\tau \frac{n_r k_B T}{3} \frac{n_{\chi}}{2 n_r + n_{\chi}}= \frac{\rho_r}{9} \tilde{\eta}\tau,
\end{equation}
 where we have introduced the parameter 
\begin{equation}\label{tileta}
    \tilde{\eta}\equiv\frac{\eta_{\chi r}}{2+\eta_{\chi r}}, \quad \eta_{\chi r} \equiv \frac{n_{\chi}}{n_r}.
\end{equation}

 While the dark matter particle number density remains unknown one can relate it to the well known quantity $n_B/n_r\simeq 6.1 \times 10^{-10}$, the baryon-to-photon ratio. The dark matter-to-photon ratio reads
\begin{eqnarray}\label{tildeetaexp}
    \eta_{\chi r}=\frac{n_{\chi}}{n_r}=\frac{n_B}{n_r}\frac{n_{\chi}}{n_B}&\approx&\frac{n_B}{n_r}\frac{\rho_{\chi}/m_{\chi}}{\rho_B/m_B}\\ \nonumber
    &\approx& 5 \frac{n_B}{n_r}\frac{m_B}{m_{\chi}}\approx \frac{2.9 }{m_{\chi}} [eV/c^2].
\end{eqnarray}

In the above estimation we have assumed that the typical baryon mass is the one of a proton and both dark matter and baryons are treated as fully non-relativistic components i.e. $k_B T/m c^2 << 1$. Also, according to the standard cosmological model, the ratio between dark matter and baryon energy densities remains constant along the entire cosmological evolution as $\rho_{\chi}/\rho_B\approx 5$ i.e., there is no particle creation process or energy density interaction between these components.

The factor $\tilde{\eta}$ in  the expression (\ref{xi+}) for the bulk-viscosity coefficient depends only on the mass $m_{\chi}$ of the dark-matter particle via (\ref{tileta})  and (\ref{tildeetaexp}). In the large-mass limit $m_{\chi}\rightarrow \infty$ both the factor $\eta_{\chi r}$ and the factor $\tilde{\eta}$ vanish and, consequently, $\xi \rightarrow 0$. On the other hand, for light dark matter candidates with masses $m_{\chi} \ll 2.9~ \mathrm{eV/c^2}$ the ratio $\eta_{\chi r}$ may become very large and, consequently, the factor $\tilde{\eta}$ approaches its maximum value, i.e., $\tilde{\eta}\rightarrow 1$. 

\begin{figure}
\centering
\includegraphics[width=\columnwidth]{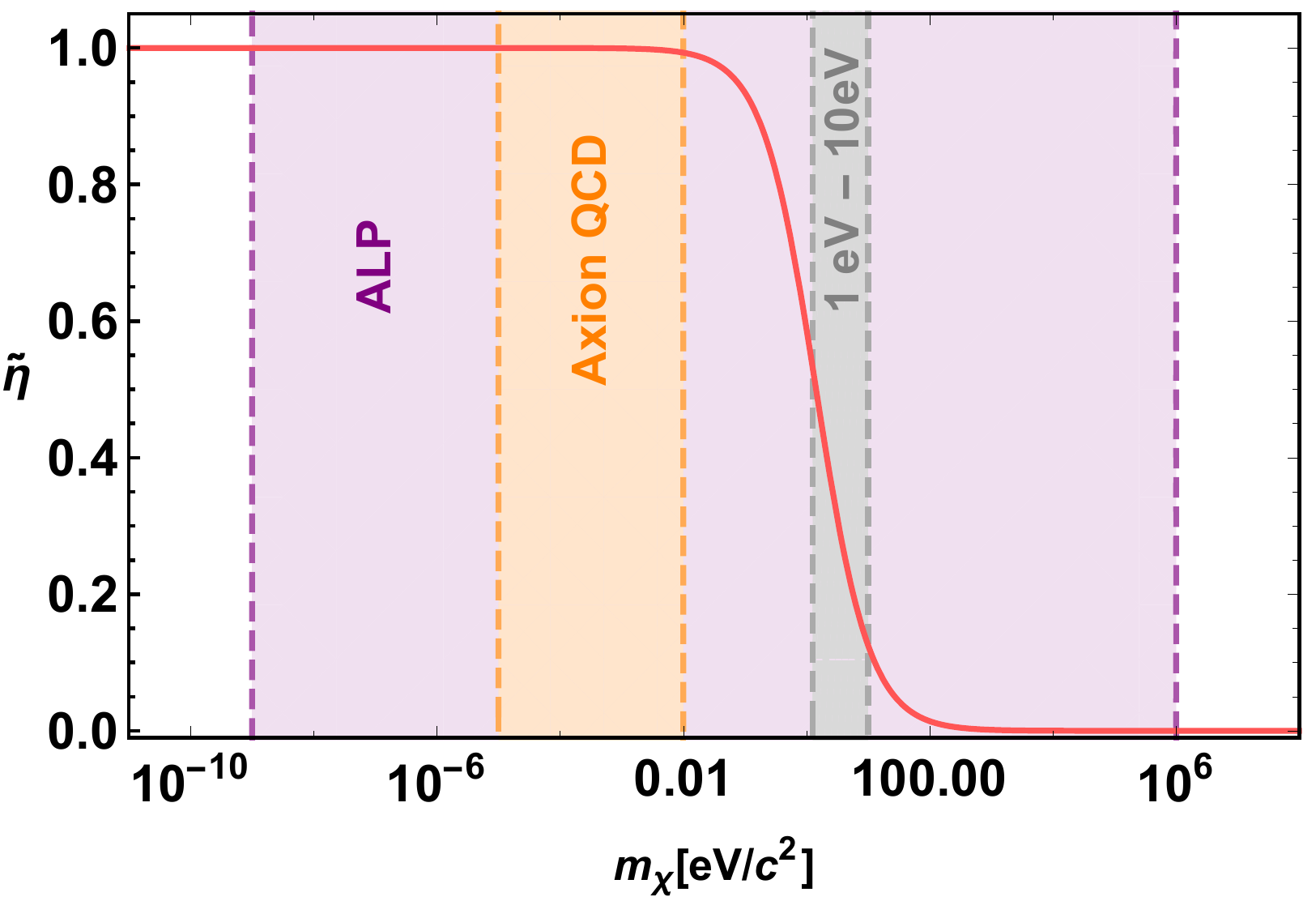}
\caption{Dependence of the factor $\tilde{\eta}$ with the dark matter particle mass $m_{\chi}$. Colored regions show the accepted mass range for Axion Like Particles (ALP) in purple and the QCD Axion in orange. The relevant mass range for this work $1eV - 10eV$ (see discussion in the text) is shown in the gray stripe.}
\label{fig:eta}
\end{figure}

Fig. \ref{fig:eta} shows quantitatively the dependence of the factor $\tilde{\eta}$ on the dark matter particle mass in ${\rm eV}$ units. 
While this effect definitely does not occur for very massive candidates like e.g., wimps with masses of ${\rm GeV}$-order, at first glance particles like axions with masses in the range $10^{-5}{\rm  eV} < m_{\rm axion} < 10^{-3} {\rm eV}$ or even lighter candidates seem to have a potentially interesting magnitude of $\xi$. However, in order to guarantee the use of (\ref{tildeetaexp}), the non-relativistic approximation in (\ref{EoS}) for the matter fluid should be valid around the equality epoch i.e., the dark matter particle should be heavier than the typical energy scale around $z_{eq}$. This sets a lower bound $m_{\chi} \gtrsim 1$eV. Therefore, a preliminary order of magnitude estimation for the desired mass range for the applicability of our approach is $1 {\rm eV}\lesssim m_{\chi} \lesssim 10 {\rm eV}$ (shown in the gray stripe of \ref{fig:eta}) which fits e.g., within the generic class of Axion Like Particles (ALP) \cite{Arias:2012az}. The ALP interpretation about the nature of our hypothetical dark matter particle guarantees that there would be no damage to LSS formation i.e., no severe cut-off on the matter power spectrum due to free streaming since such axionic mass scale corresponds effectively to a CDM particle \cite{Marsh:2015xka}.

The bulk viscosity $\xi$ depends directly on the so far unspecified time scale $\tau$. As a macroscopic scale, $\tau$ is expected to be at least slightly larger than the mean free time of the underlying microscopic dynamics. 
Generally, a fluid description of the cosmic medium is valid as long as this mean free time is much smaller than the Hubble time $H^{-1}$. 
For the scale $\tau$ we require $\tau \ll H^{-1}$ as well. 
A perfect fluid description, equivalent to local equilibrium, is valid if this scale is negligible compared with $H^{-1}$. A dissipative effect comes into play if first-order deviations from local equilibrium have to be taken into account. Our model explores the idea that a slight deviation from local equilibrium might be relevant around the epoch of matter-radiation equality. Both well before and well after the equality epoch, perfect-fluid descriptions are assumed to be valid. 

In order to describe the cosmic evolution around the equality time it is  convenient to define the variable $y=a/a_{\rm eq}=\rho_m / \rho_r$ where the subindex $m$ refers to the sum of baryons and dark matter. Hence, given the desired behavior, i.e., a short non-equilibrium period,  
discussed above,  
a convenient parameterization for the  time scale $\tau$ in (\ref{xi+})  is
\begin{equation}\label{tau}
\tau(y)=\tau_{\rm eq}\frac{H_{\rm eq}}{H}\left(\frac{2 y^2}{1+y^{4}}\right).
\end{equation}
The subindex $_{\rm eq}$ refers to the value that quantities have at the equality time e.g., $\tau_{\rm eq}= \tau(y=1)$. With this definition the time dependence of $\tau$ is modelled by only one new phenomenological parameter $\tau_{\rm eq}$ which also represents the maximal value $\tau$ can take.

The transient coupling between dark matter and radiation employed here does not rely on a specific microscopic interaction model. This would require the specification of extra physical parameters beyond the dark matter particle mass. In fact, the interaction becomes effective only when both fluids have similar contributions to the total energy density. This is a consequence of the thermodynamical description employed previously. Both earlier and later on, when one of the components fully dominates, the interaction vanishes. The parameterization proposed in (20) captures this phenomenology.

The fluid description employed here is valid as long as $\tau << H^{-1}$ which is guaranteed for $\tau_{\rm eq} H_{\rm eq}<< 1$. 

\begin{figure}
\centering
\includegraphics[width=\columnwidth]{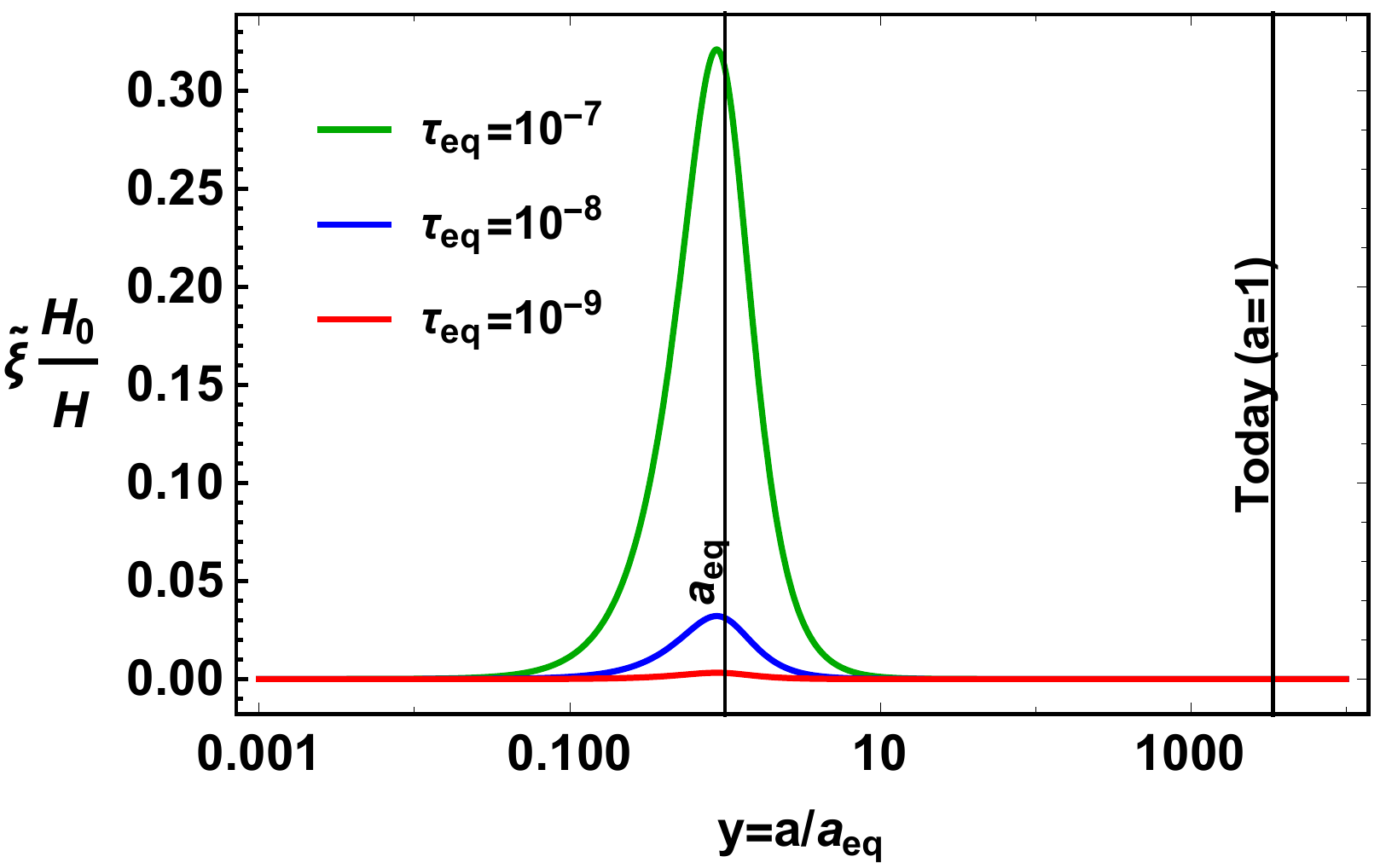}
\caption{Effective bulk viscosity as a function of the variable $y = a / a_{\rm eq}$ for various values of $\tau_{\rm eq}$ (in seconds). Matter-radiation equality occurs at $y=1$ (first vertical solid line) corresponding to the redshift $z_{\rm eq}=a_{\rm eq}^{-1}-1=3402$. The second vertical solid black line indicates the today's scale factor $a=1$.}
\label{fig:tildexi}
\end{figure}

In order to estimate the magnitude of the bulk viscous coefficient $\xi$ we adopt Planck 2018 cosmological parameters in which $\Omega_{m0}=8\pi G \rho_{m0}/3 H^2_0=0.315$ and the redshift of equality is $z_{\rm eq}=3402$ \cite{Aghanim:2018eyx}. As we shall shown below, the effective quantity entering both the background dynamics as well as the equation for the growth of linear scalar perturbations, is the dimensionless combination 
\begin{equation}\label{fracPi}
\frac{3\Pi}{\rho} = - 24 \pi G H^{-1} \xi \equiv -
\tilde{\xi} \frac{H_0}{H}. 
\end{equation}
With this definition Eq. (\ref{xi+}) is written as  
\begin{equation}\label{tildexi}
    \tilde{\xi}\frac{H_0}{H} =\tau(y) \frac{H^2_0}{H} \Omega_r \, \tilde{\eta},
\end{equation}
where $\Omega_r=\rho_r/\rho_{0}=8 \pi G \rho_r/3 H^2_0$. Hence, in Fig. \ref{fig:tildexi} we show the evolution of this quantity as a function of the $y$ variable for different values of the $\tau_{\rm eq}$ parameter. Here and henceforth in this work we adopt $m_{\chi}=1$eV which corresponds to $\tilde{\eta}\simeq 0.59$.

As expected, the bulk viscosity vanishes both deep in the radiation epoch ($y<<1$) and later on during pure matter domination $(y>>1)$. However, it has a non-vanishing contribution to the total pressure around matter-radiation equality.

Notice that since $H_{\rm eq}$ is of the order of $10^7 {\rm s}^{-1} $, the case 
$\tau_{\rm eq} = 10^{-7} {\rm s}$ marks the applicability limit of our approach.

\section{Background expansion and the $H_0$ tension}

In this section we explore the impact of a non-vanishing bulk viscosity of the described type on the background expansion. 
While close to equilibrium the number and energy densities coincide with with the local equilibrium values, the pressure does not. 
The expression (\ref{tildexi})  measures the deviation from the equilibrium pressure. It gives rise to a modification of the effective equation of state of the cosmic substratum during the period in which $\tilde{\xi} \neq 0$. 
For earlier and later times the standard expansion behavior is recovered.

We assume that before and after the period with $\tilde{\xi} \neq 0$ the Universe evolves according to the standard flat-$\Lambda$CDM model with ($\tilde{\xi} = 0$)
\begin{equation}\label{ExpansionLCDM}
    \frac{H^2_{\Lambda}(a)}{H_0^2}=\Omega_m (a)+\Omega_r(a)+\Omega_{\Lambda}, 
\end{equation}
where $\Omega_m=\rho_m/\rho_{0}$ and $\Omega_{\Lambda}=1-\Omega_{m0}-\Omega_{r0}$. Expression (\ref{ExpansionLCDM}) will be used to set the initial conditions for the evolution of the Hubble rate. 

Now, let us find how the expansion rate changes when there exists an extra bulk viscous pressure contribution (\ref{fracPi}). 
We start considering the conservation balance for total energy density $\rho=\rho_m + \rho _r + \rho_{\Lambda}$ equipped with a total pressure $p=p_r+p_{\Lambda}+\Pi$. This reads
\begin{equation}
    \dot{\rho}+3H(\rho+p_r+p_{\Lambda}+\Pi)=0.
\label{contTOTAL}\end{equation}
With the help of the Friedmann equation for the total energy density $\rho=3 H^2/8\pi G$ and defining the dimensionless expansion parameter $E=H/H_0$ we can rewrite Eq. (\ref{contTOTAL}) as
\begin{equation}\label{Omega}
    2 E a \frac{dE}{da} + 3E^2\left(1-{\frac{\tilde{\xi}}{3E}}\right) +\frac{\Omega_{r0}}{a^4} - 3 (1-\Omega_{r0}-\Omega_{m0})=0.
\end{equation}
For $\tilde{\xi}=0$ the standard  $\Lambda$CDM cosmology (\ref{ExpansionLCDM}) is recovered.

 In order to assess the impact of the bulk viscous contribution to the background expansion one has to solve (\ref{Omega}) with values $\tilde{\xi}>0$. The expansion rate (\ref{ExpansionLCDM}) is used to set the initial expansion deep in the radiation dominated epoch at an initial scale factor $a_i$, e.g., $a_i=a_{\rm eq}/10000$. By evolving (\ref{Omega}) numerically with initial condition 
\begin{equation}
    E (a_i)=\frac{H_{\Lambda}(a_i)}{H^{cmb}_{0}},
\end{equation} 
where we adopt $H_0^{cmb}=67.4 \,km\, s^{-1} \,Mpc^{-1}$ \cite{Aghanim:2018eyx},
 until the present time with scale factor $a_0=1$, we obtain the corresponding present Hubble rate, influenced by the bulk viscous contribution around the matter-radiation equality epoch.
In the absence of a bulk viscous pressure, i.e., for $\tilde{\xi}=0$, the ratio
 $E$ is normalized to $E^{cmb}(a_0 =1) = 1$ with $H_0 = H_0^{cmb}$. If there is a period during which the bulk viscous pressure becomes dynamically relevant, Eq. (\ref{Omega}) with the same initial condition will result in $E(a_0 =1) \neq 1$ which corresponds to a value $H_0 = E(a_0 =1)H_0^{cmb}$.

\begin{figure}
\centering
\includegraphics[width=\columnwidth]{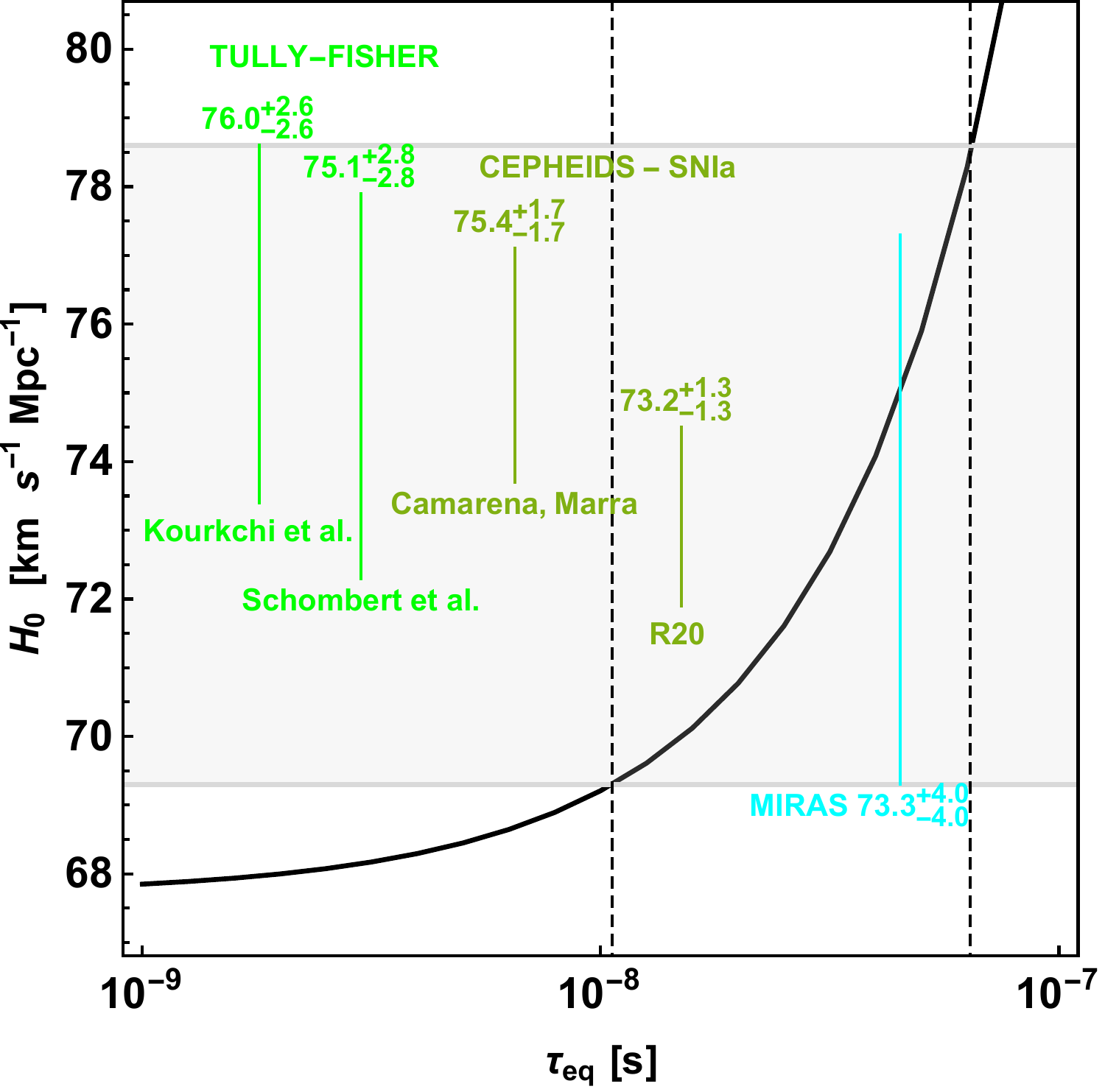}
\caption{The black curve shows the dependence of $H_0$ on the $\tau_{\rm eq}$ value. Some relevant constraints of the Hubble constant $H_0$ through direct measurement methods are shown. Among them, the measurements based on the Tully-Fisher method as in Kourchi et al. \cite{tullyfisher1} and  Schombert et al. \cite{tullyfisher2} provide the largest $H_0$ values (including uncertainties). 
We include the latest SH0ES result, based on direct Cepheid-SNIa relations ((R20) \cite{sh0es}).
By changing the marginalization process over free parameters, Camarena \& Marra have obtained a higher $H_0$ value from available Cepheid-SNIa data \cite{camarena}. We also include the MIRAS result \cite{miras} which provides one of the lowest $H_0$ values (including uncertainties) among the direct distance ladder measurements of $H_0$. The range between the highest and the lowest $H_0$ values from direct methods determines the gray filled region which is used to set the range of acceptable $\tau_{\rm eq}$ values that alleviate the $H_0$ tension. Vertical dashed lines set the boundaries $ 1.06\times 10^{-8} \lesssim \tau_{\rm eq}\lesssim 6.4 \times 10^{-8}$.}
\label{fig:H}
\end{figure}

We show in Fig. \ref{fig:H} the $H_0$ dependence on the $\tau_{\rm eq}$ parameter value. In the vanishing viscosity limit $\tau_{\rm eq} \rightarrow 0$ the black line tends to $H_0=67.4 \,km\, s^{-1} \,Mpc^{-1}$ as expected. The gray horizontal stripe covers the available range of distance ladder measurements for $H_0$ including uncertainties. For $\tau_{\rm eq}$ values in the range $ 1.06\times 10^{-8} \lesssim \tau_{\rm eq}\lesssim 6.4 \times 10^{-8}$ the today's expansion rate fits within the measured range of $H_0$ values from distance ladder probes.

\section{Perturbations}
Now we explore the consequences of the dynamics 
outlined  in the last section for large-scale structure formation.


\subsection{The Mészáros effect}

We start by reviewing the Mészáros equation which describes the evolution of fractional non-relativistic matter perturbations $\delta_m \equiv \frac{\delta\rho_m}{\rho_m}$  in a radiation background.  The standard equation for $\delta_m$, 
\begin{equation}
    \ddot{\delta}_m+2H \dot{\delta}_m-4\pi G \rho_m \delta_m=0, 
\label{eqdelta}
\end{equation}
is obtained by combining first order versions of the continuity, Euler and Poisson equations.



The Hubble expansion is influenced by the effective equation of state $w$ of the cosmic medium via 
\begin{equation}
\dot{H}=- 4 \pi G \rho (1+w). 
\end{equation}

Eq. (\ref{eqdelta}) can be rewritten in terms of the $y-$variable such that the evolution of matter overdensity becomes the Mészáros equation \cite{Meszaros:1974tb}

\begin{equation}
    \frac{d ^2\delta_m}{dy^2}+\left(\frac{1}{H}\frac{d H}{dy}+\frac{3}{y}\right)\frac{d \delta_m}{dy}-\frac{3}{2y\left(1+y\right)}\delta_m=0.
\end{equation}

The effect of the background expansion is  encoded in the function 
\begin{equation}
    \frac{1}{H}\frac{d H}{dy}=-\frac{3}{2}\frac{(1+w)}{y}.
\end{equation}

Specifying to the standard equations of state for a mixture of non-relativistic matter and radiation, 
 Eq. (\ref{eqdelta}), in terms of the $y$- variable, becomes  
\begin{equation}
\label{d2delta}
    \frac{d ^2\delta_m}{dy^2}+\frac{\left(2+3y\right)}{2y \left(1+y\right)}\frac{d \delta_m}{dy}-\frac{3}{2y\left(1+y\right)}\delta_m=0.
\end{equation}
The above equation has an analytical growing mode solution of the type $\delta_m \sim y + 2/3$. Deep in the radiation epoch ($y<<1$) the quantity $\delta$ remains constant while it grows linearly with the scale factor in the matter dominated period ($y>>1$).

We explore now how the emergence of a transient effective bulk viscosity at background cosmological level impacts the evolution of dark matter perturbations through matter-radiation equality. 
Strictly speaking, the bulk viscous pressure itself can be split into background and first-order parts, the latter accompanied by a scale dependence   proportional to $k^2$ where $k$ is the perturbation wavenumber \cite{HipolitoRicaldi:2009je,HipolitoRicaldi:2010mf,Velten:2011bg,Velten:2012uv,Velten:2013pra,Velten:2014xca,Velten:2015tya,Barbosa:2017ojt}. Here, we focus on the background contribution only.



Taking into account a bulk viscous pressure contribution (\ref{fracPi}), equation (\ref{d2delta}) is modified to yield
\begin{equation}\label{delta}
    \frac{d ^2\delta_m}{dy^2}+\left[\frac{\left(2+3y\right)}{2y \left(1+y\right)}+\frac{\tilde{\xi}}{2y}\frac{H_0}{H}\right]\frac{d \delta_m}{dy}-\frac{3}{2y\left(1+y\right)}\delta_m=0.
\end{equation}

Since the bulk viscous coefficient should obey  $\xi>0$, the total Hubble friction is enhanced leading to a growth suppression. The magnitude of the viscous quantity $\tilde{\xi} H/H_0$ in the Hubble friction term is shown in Fig. \ref{fig:tildexi} for different  $\tau_{{\rm eq}}$ values.

\begin{figure}
\centering
\includegraphics[width=\columnwidth]{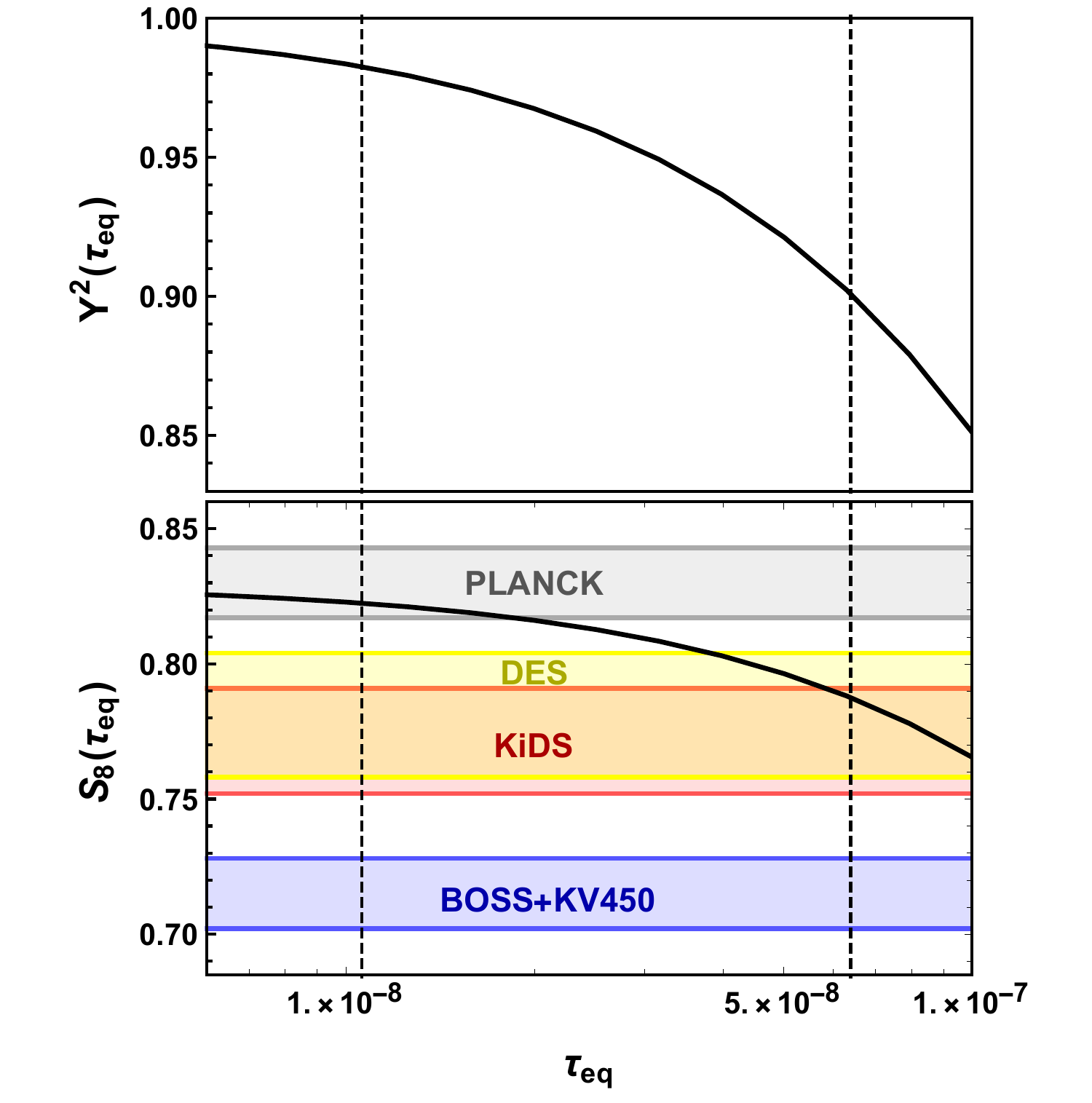}
\caption{The new transfer function ${\rm Y}^2$ of the matter power spectrum (top panel) and the $S_8$ quantity (lower panel) as a function of the interaction time scale $\tau_{\rm eq}$. Vertical dashed lines set the boundaries $1.06 \times 10^{-8} \lesssim \tau_{\rm eq}\lesssim 6.4 \times 10^{-8}$ as obtained from Fig. \ref{fig:H}.}
\label{fig:YS8}
\end{figure}

The matter power spectrum $P(k)=\left|\delta_{k}\right|^2$ is defined as the mean square amplitude of the Fourier components of the perturbed density field. The primordial spectrum set at the end of inflation with a power law shape $P_{i}=A k^{n}$, where $A$ is the initial amplitude,  has its spectral index $n$ constrained by observations to $n=0.96$. The today's observed power spectrum evolves from $P_{i}$ by taking into account the evolution of linear matter perturbations. This effect is encoded in the scale independent growth function $D_{+}=\delta(t)/\delta(t=t_0)$. Also, the growth suppression experienced by $k-$modes that enter the horizon at the radiation dominated epoch leads to the typical curved shape seen in the power spectrum. The smaller the scale, the larger the growth suppression. The  scale-dependence due to this process is captured by the transfer function $T(k)$.  In order to describe the additional   suppression due to  viscous background effects around $z_{\rm eq}$ we define a $\tau_{\rm eq}$ dependent transfer function $Y(\tau_{\rm eq})$ with 
$Y^2(\tau_{\rm eq} = 0) = 1$, such that the observed today's matter power spectrum $P(k)$,  calculated from (\ref{delta}), is written as
\begin{equation}
P(k)={\rm Y}^2(\tau_{\rm eq}) D_{+}^2\,  T^2(k) A k^n. 
\end{equation}
In the absence of viscous effects, $Y^2 = 1$ is valid and the spectrum coincides with the corresponding quantity calculated from (\ref{d2delta}) with the same initial conditions.
The total growth suppression encoded in ${\rm Y^2}(\tau_{\rm eq})$ is plotted in the upper panel of Fig. \ref{fig:YS8}. For $\tau_{\rm eq}$ values of order $10^{-8}$ a $10\%$ effect on $P(k)$ is seen which is compatible with the uncertainty level present in current $P(k)$ measurements \cite{Gil-Marin:2016wya}.

There also exists a relevant $\sim 2 \sigma$ tension between the predicted amplitude of matter clustering parameter
\begin{equation}
    S_8=\sigma_8 \left(\frac{\Omega_{m0}}{0.3}\right)^{1/2}
\end{equation}
based on the Planck parameters cosmology and its measurement in the local universe. While Planck has obtained $S_8=0.830\pm 0.013$ local measurements have found $S_8$ values slightly smaller: $S_8= 0.766^{+0.020}_{-0.014}$ (KiDS-1000 \cite{heymans}), $S_8= 0.783^{+0.021}_{-0.025}$ (DES \cite{des1, des2}), $S_8=0.728\pm 0.026$ (BOSS+KV450 \cite{troster}).

As seen in the lower panel of Fig. \ref{fig:YS8}, values $\tau_{\rm eq} \approx 10^{-8}$ are not yet able to fully address the $S_8$ tension.
While $\tau_{\rm eq} \approx 7 \times 10^{-8}$ agrees with $KiDS$ and $DES$ data, it is outside the range of $BOSS+KV450$ data.

\subsection{Scales larger than the horizon}
Super-horizon scales can be assessed via relativistic cosmological perturbation theory. By using the Newtonian gauge the perturbed metric for scalar perturbations reads
\begin{equation}
    ds^2=a^2\left[-(1+2\Phi)d\eta^2 + (1+2\Psi) \delta_{ij}dx^i dx^j\right].
\end{equation}
In the absence of anisotropic stresses $\Phi=\Psi$ is valid. With this condition the $0-0$ and the $i=j$ components of the Einstein equations read, respectively,
\begin{equation}
    3\mathcal{H}\left(\Phi^{\prime}+\mathcal{H}\Phi\right)-\nabla^2\Phi=-4\pi G a^2 \delta\rho,
\end{equation}
and 
\begin{equation}\label{eqpot}
    \Phi^{\prime\prime}+3\mathcal{H}\Phi^{\prime}+\left(2 \mathcal{H}^{\prime} +\mathcal{H}^2\right)\Phi=4\pi G a^2 \delta P,
\end{equation}
where a prime ``$\,^{\prime}\,$'' means derivative with respect to the conformal time. Scale dependence (in the Fourier space) enters into the above equations via  $\nabla^2\rightarrow -k^2$.

In the standard cosmology the pressure perturbation $\delta P$ is identified with the total density perturbation via the effective speed of sound $c^2_{eff}= \delta P / \delta \rho$. Here, the bulk viscous pressure perturbation is added to the radiation pressure perturbation i.e.,
\begin{equation}
    \delta P= \delta p_r + \delta \Pi.
\end{equation}
Writing the above quantities in a covariant way, as expected within the relativistic formalism, the expansion scalar $\Theta$ up to first order in cosmological perturbations reads

\begin{equation}
    \Theta = \frac{3\mathcal{H}}{a}-\frac{3\mathcal{H}\Phi}{a}-\frac{3\Psi^{\prime}}{a}+\delta u^{i}_{,i},
\end{equation}
where $\delta u^{i}_{,i} \equiv -kv/a$ where $v$ is the four-velocity potential. But this term will not be relevant since for super-horizon scales $k<<\mathcal{H}$ the scale-dependent terms are neglected. Again, in terms of the $y-$variable Eq. (\ref{eqpot}) becomes


\begin{eqnarray}
    &&\frac{d^2 \Phi}{dy^2}+\left[\frac{21y^2+54y+32}{2y(1+y)(3y+4)}\right]\frac{d\Phi}{dy}\nonumber \\
    &+&\frac{\Phi}{y(1+y)(3y+4)}=\frac{\tilde{\xi}}{2 y^2}\frac{H_0}{H}\Phi+\frac{\tilde{\xi}}{2y}\frac{H_0}{H}\frac{d \Phi}{dy}.
\end{eqnarray}


\begin{figure}[t!]
\centering
\includegraphics[width=\columnwidth]{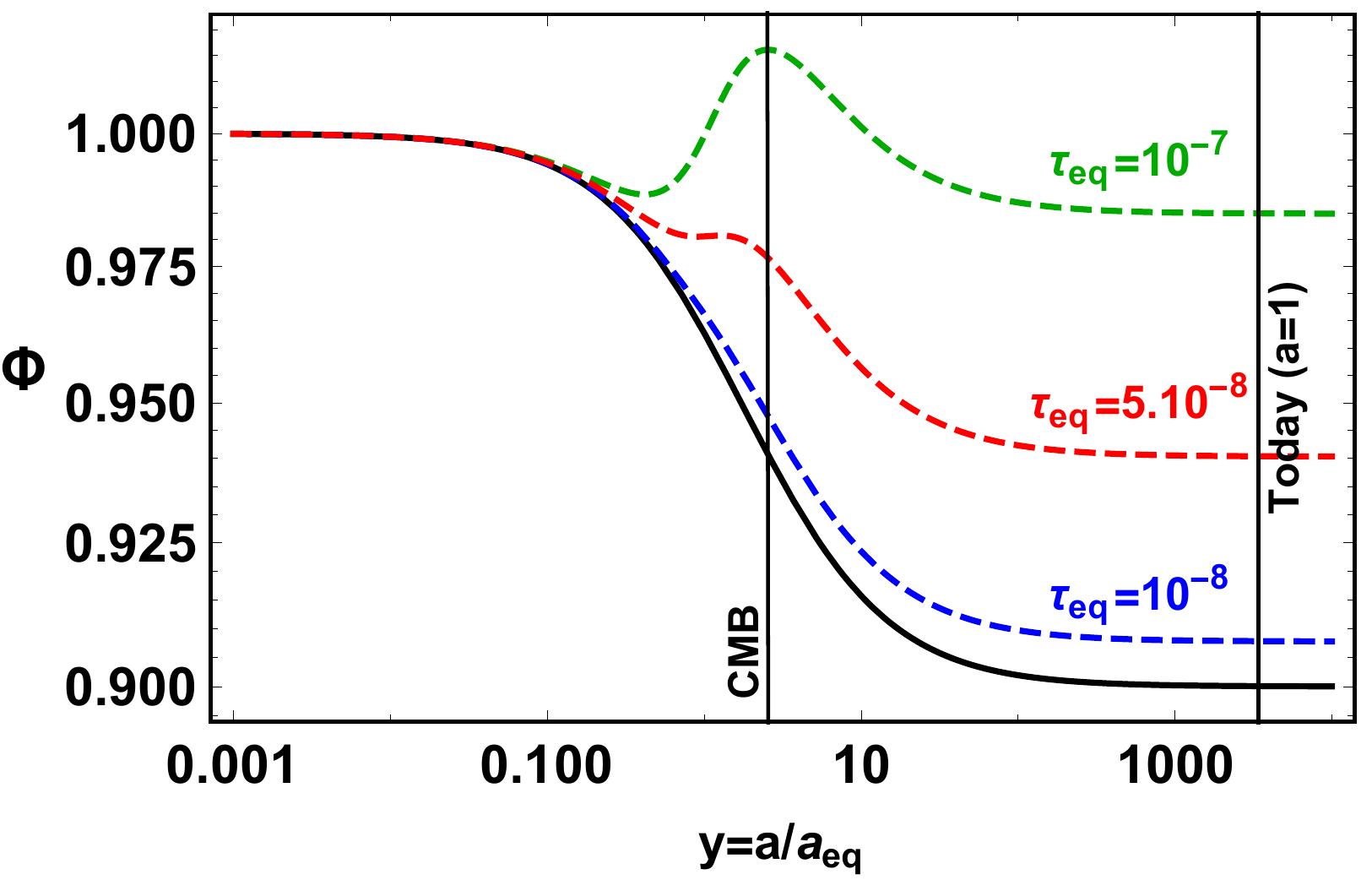}
\caption{Evolution of the large scale gravitational potential $\Phi$ as a function of the $y$ variable. Vertical solid lines denote the CMB epoch ($z_{cmb} \approx 1100$) and today ($a_0=1$). Solid curve shows the standard flat $\Lambda$CDM behavior in which the potential decays to $9/10$ of its initial value. The  green (red) [blue] dashed curves show how the potential evolves for $\tau_{\rm eq}=10^{-7} \, (\tau_{\rm eq}= 5 \times 10^{-8})[\tau_{\rm eq}=10^{-8}]$.}
\label{fig:3}
\end{figure}

In the radiation dominated era the solution of this equation  reduces to a constant initial amplitude $\Phi_{i}$. Once the Universe becomes matter dominated the gravitational potential solution gives $\Phi \rightarrow (9/10) \Phi_{i}$ for $y>>1$. Then, perturbations that enter the horizon after the epoch of matter–radiation equality have their amplitudes reduced by a factor of 1/10 through the radiation–matter equality epoch. Fig. \ref{fig:3}  shows the dependence of $\Phi_{i}$ on $y$ for various values of $\tau_{\rm eq}$. As to be seen, for $\tau_{\rm eq}\sim 10^{-8}$ we expect a mild $\sim 2\%$ impact on the large scale potential at late times while the impact on the CMB epoch should be even smaller. On the other hand, we can surely rule out values $\tau_{\rm eq} \sim 10^{-7}$ since for them  the gravitational potential is amplified along the transition.

\section{Conclusions}

A perfect fluid description of cosmic fluids is one of the theoretical pillars of the standard cosmological model. In a two-fluid model of the Universe an interaction between both components may result in a transient close-to-equilibrium state. As a result, the cosmic substratum as a whole acquires an effective bulk viscous pressure. We have explored the consequences of such phenomenon for a mixture of radiation and matter around the epoch of radiation-matter equality at a redshift of the order of $z \approx 3400$. For the specific value of the bulk viscosity  we have used a simple gas model which allows for an analytic calculation, based on Eckart's theory. We expect, however, the general aspects of our approach to be valid for bulk-viscous pressures of different origin as well.



The relevance of this effect in our context depends on the mass of the dark-matter particle.
The fluid description employed here imposes $1 {\rm eV} \lesssim m_{\chi} \lesssim 10  {\rm eV}$ as the optimal range for the appearance of a transient bulk viscous pressure at the background level. By fixing $m_{\chi}=1 {\rm eV}$, this mechanism can provide a hint for solving the current cosmic tensions associated to the $H_0$ measurements and the matter clustering features for a characteristic relaxation time interval $\tau_{\rm eq} \sim 7 \times 10^{-8}$
at the matter-radiation epoch. Such value is consistent with the validity of the fluid formalism of our approach.

We provide therefore a possible hint for solving the current cosmic tensions by adding a new ingredient to the cosmological description at early times. Indeed, it has been recently argued that changes to late time physics can not be seen as an appropriate way to solve the $H_0$ tension \cite{Efstathiou:2021ocp}.

The phenomenology explored in this work should be further studied with a detailed quantitative statistical analysis using mainly CMB data. From the results presented here we expect an extra imprint on the early integrated Sachs-Wolfe effect which measures the time variation of the gravitational potential just after the CMB photons decoupling. This will be the subject of a future work.

{\bf Acknowledgments}
WZ acknowledges Conselho Nacional de Desenvolvimento Científico e Tecnológico (CNPq) for financial support under project APV 451916/2019-0. HV thanks CNPq/FAPES/CAPES and PROPP/UFOP for partial financial support. IC thanks CNPQ for financial support. We thank David Camarena for useful discussion.


\begin{thebibliography}{99}

\bibitem{Meszaros:1974tb}
P.~Meszaros,
Astron. Astrophys. \textbf{37}, 225-228 (1974).


\bibitem{amendola}
L. Amendola and S. Tsujikawa, { \it Dark Energy: Theory and Observations}, Cambridge, Cambridge University Press (2010).

\bibitem{mo} 
H. Mo, F. van den Bosch and S. White, {\it  
Galaxy Formation and Evolution}, Cambridge, Cambridge University Press (2010).


\bibitem{Silk} J. Silk, ApJ, {\bf151}, 459 (1968).

\bibitem{Pan:2018zha}
Z.~Pan, M.~Kaplinghat and L.~Knox,
Phys. Rev. D \textbf{97} (2018) no.10, 103531
doi:10.1103/PhysRevD.97.103531
[arXiv:1801.07348 [astro-ph.CO]].
\bibitem{Das:2017nub}
S.~Das, R.~Mondal, V.~Rentala and S.~Suresh,
JCAP \textbf{08} (2018), 045
doi:10.1088/1475-7516/2018/08/045
[arXiv:1712.03976 [astro-ph.CO]].

\bibitem{Zimdahl:1996fj}
W.~Zimdahl,
Mon. Not. Roy. Astron. Soc. \textbf{280} (1996), 1239.

\bibitem{Arias:2012az}
P.~Arias, D.~Cadamuro, M.~Goodsell, J.~Jaeckel, J.~Redondo and A.~Ringwald,
JCAP \textbf{06} (2012), 013
doi:10.1088/1475-7516/2012/06/013
[arXiv:1201.5902 [hep-ph]].


\bibitem{Marsh:2015xka}
D.~J.~E.~Marsh,
Phys. Rept. \textbf{643}, 1-79 (2016)
doi:10.1016/j.physrep.2016.06.005
[arXiv:1510.07633 [astro-ph.CO]].

\bibitem{Zimdahl:1996ka}
W.~Zimdahl,
Phys. Rev. D \textbf{53} (1996), 5483-5493
doi:10.1103/PhysRevD.53.5483
[arXiv:astro-ph/9601189 [astro-ph]].

\bibitem{Colistete:2007xi}
R.~Colistete, J.~C.~Fabris, J.~Tossa and W.~Zimdahl,
Phys. Rev. D \textbf{76} (2007), 103516
doi:10.1103/PhysRevD.76.103516
[arXiv:0706.4086 [astro-ph]].

\bibitem{Brevik:2020psp}
I.~Brevik and B.~D.~Normann,
Symmetry \textbf{12} (2020) no.7, 1085
doi:10.3390/sym12071085
[arXiv:2006.09514 [gr-qc]].

\bibitem{Normann:2016jns}
B.~D.~Normann and I.~Brevik,
Entropy \textbf{18} (2016), 215
doi:10.3390/e18060215
[arXiv:1601.04519 [gr-qc]].

\bibitem{Szydlowski:2020awp}
M.~Szyd\l{}owski and A.~Krawiec,
Symmetry \textbf{12} (2020) no.8, 1269
doi:10.3390/sym12081269

\bibitem{Brevik:2017msy}
I.~Brevik, \O{}.~Gr\o{}n, J.~de Haro, S.~D.~Odintsov and E.~N.~Saridakis,
Int. J. Mod. Phys. D \textbf{26} (2017) no.14, 1730024
doi:10.1142/S0218271817300245
[arXiv:1706.02543 [gr-qc]].

\bibitem{Barbosa:2015ndx}
C.~M.~S.~Barbosa, J.~C.~Fabris, O.~F.~Piattella, H.~E.~S.~Velten and W.~Zimdahl,
[arXiv:1512.00921 [astro-ph.CO]].





\bibitem{Udey}
N. Udey and W. Israel, Mon. Not. Roy. Astron. Soc. \textbf{119} (1982), 1137.




\bibitem{Eckart}
C. Eckart,The Thermodynamics of irreversible processes. 3.. Relativistic theory of the simple fluid, Phys.Rev.58(1940) 919–924



\bibitem{Aghanim:2018eyx}
N.~Aghanim \textit{et al.} [Planck],
[arXiv:1807.06209 [astro-ph.CO]].


\bibitem{HipolitoRicaldi:2009je}
W.~S.~Hipolito-Ricaldi, H.~E.~S.~Velten and W.~Zimdahl,
JCAP \textbf{06} (2009), 016
doi:10.1088/1475-7516/2009/06/016
[arXiv:0902.4710 [astro-ph.CO]].

\bibitem{HipolitoRicaldi:2010mf}
W.~S.~Hipolito-Ricaldi, H.~E.~S.~Velten and W.~Zimdahl,
Phys. Rev. D \textbf{82} (2010), 063507
doi:10.1103/PhysRevD.82.063507
[arXiv:1007.0675 [astro-ph.CO]].

\bibitem{Velten:2011bg}
H.~Velten and D.~J.~Schwarz,
JCAP \textbf{09} (2011), 016
doi:10.1088/1475-7516/2011/09/016
[arXiv:1107.1143 [astro-ph.CO]].


\bibitem{Velten:2012uv}
H.~Velten and D.~Schwarz,
Phys. Rev. D \textbf{86} (2012), 083501
doi:10.1103/PhysRevD.86.083501
[arXiv:1206.0986 [astro-ph.CO]].


\bibitem{Velten:2013pra}
H.~Velten, D.~Schwarz, J.~Fabris and W.~Zimdahl,
Phys. Rev. D \textbf{88} (2013) no.10, 103522
doi:10.1103/PhysRevD.88.103522
[arXiv:1307.6536 [astro-ph.CO]].

\bibitem{Velten:2014xca}
H.~Velten, T.~R.~P.~Caramês, J.~C.~Fabris, L.~Casarini and R.~C.~Batista,
Phys. Rev. D \textbf{90} (2014) no.12, 123526
doi:10.1103/PhysRevD.90.123526
[arXiv:1410.3066 [astro-ph.CO]].

\bibitem{Velten:2015tya}
H.~Velten,
AIP Conf. Proc. \textbf{1647} (2015) no.1, 76-79
doi:10.1063/1.4913342

\bibitem{Barbosa:2017ojt}
C.~M.~S.~Barbosa, H.~Velten, J.~C.~Fabris and R.~O.~Ramos,
Phys. Rev. D \textbf{96} (2017) no.2, 023527
doi:10.1103/PhysRevD.96.023527
[arXiv:1702.07040 [astro-ph.CO]].


\bibitem{tullyfisher1}
E. Kourkchi, R. B. Tully, G. S. Anand, H. M. Courtois, A. Dupuy, J. D. Neill, L. Rizzi and M. Seibert, Astrophys. J. \textbf{896} (2020) 3 [arXiv:2004.14499 [astro-ph.CO]]


\bibitem{tullyfisher2}
J. Schombert, S. McGaugh and F. Lelli, Astron. J. \textbf{160} (2020) 71 [arXiv:2006.08615 [astro-ph.CO]]

\bibitem{sh0es}
A. G. Riess, S. Casertano, W. Yuan, J. B. Bowers, L. Macri, J. C. Zinn and D. Scolnic, Astrophys. J. Lett. \textbf{908} (2021) L6 [arXiv:2012.08534 [astro-ph.CO]]


\bibitem{camarena}
D. Camarena and V. Marra, Phys. Rev. Res. \textbf{2} (2020) 013028 [arXiv:1906.11814  [astro-ph.CO]]




\bibitem{miras}
C. D. Huang, A. G. Riess, W. Yuan, L. M. Macri, N. L. Zakamska, S. Casertano, P. A. Whitelock, S. L. Hoffmann, A. V. Filippenko and D. Scolnic 2019 [arXiv:1908.10883 [astro-ph.CO]]


\bibitem{Gil-Marin:2016wya}
H.~Gil-Marín, W.~J.~Percival, L.~Verde, J.~R.~Brownstein, C.~H.~Chuang, F.~S.~Kitaura, S.~A.~Rodríguez-Torres and M.~D.~Olmstead,
Mon. Not. Roy. Astron. Soc. \textbf{465} (2017) no.2, 1757-1788
doi:10.1093/mnras/stw2679
[arXiv:1606.00439 [astro-ph.CO]].


\bibitem{heymans}
C. Heymans  \textit{et al}.,
7, 2020
[arXiv:2007.15632 [astro-ph.CO]].



\bibitem{des1}
\textbf{DES} Collaboration, T. M. C. Abbott \textit{et al}., 
\textit{ Phys. Rev.} \textbf{D98} no. 4, (2018) 043526
doi:10.1103/PhysRevD.98.043526
[arXiv:1708.01530 [astro-ph.CO]].


\bibitem{des2}
\textbf{DES} Collaboration, M. A. Troxel \textit{et al}.,
\textit{ Phys. Rev.} \textbf{D98} no. 4, (2018) 043528
doi:10.1103/PhysRevD.98.043528
[arXiv:1708.01538 [astro-ph.CO]].




\bibitem{troster}
T. Tröster \textit{et al}., 
\textit{Astron. Astrophys.} \textbf{633} (2020) L10 
doi:10.1051/0004-6361/201936772 
[arXiv:1909.11006 [astro-ph.CO]].


\bibitem{Efstathiou:2021ocp}
G.~Efstathiou,
[arXiv:2103.08723 [astro-ph.CO]].

\end{thebibliography}
\end{document}